\documentclass[a4paper]{siamart0216}
\usepackage{amsmath}
\usepackage{amsfonts}
\usepackage{amssymb}
\usepackage{amsopn}
\usepackage{mathrsfs}
\usepackage{graphicx}
\usepackage{subfig}
\usepackage{pstool}
\usepackage{epstopdf}
\usepackage{verbatim}
\usepackage{algorithmic}
\usepackage{tikz}
\usetikzlibrary{shapes,arrows}

\newcommand{\eg}{\emph{e.g. }}
\newcommand{\ie}{\emph{i.e. }}
\newcommand{\pare}[1]{\left( #1 \right)}
\newcommand{\emE}{\ensuremath{\vec{E}}}
\newcommand{\emD}{\ensuremath{\vec{D}}}
\newcommand{\emB}{\ensuremath{\vec{B}}}
\newcommand{\emH}{\ensuremath{\vec{H}}}

\newcommand{\e}{\ensuremath{\varepsilon}}
\newcommand{\bare}{\ensuremath{\bar{\e}}}
\newcommand{\barm}{\ensuremath{\bar{\mu}}}
\newcommand{\kap}{\ensuremath{\bar{\kappa}}}
\newcommand{\eo}{\ensuremath{\e_o}}
\newcommand{\mo}{\ensuremath{\mu_o}}

\newcommand{\q}{\ensuremath{\vec{q}}}
\newcommand{\fq}{\ensuremath{\vec{f}\left(\q\right)}}
\newcommand{\f}{\vec{f}}
\newcommand{\W}{\mathcal{W}}
\newcommand{\A}{\mathcal{A}}
\newcommand{\B}{\mathcal{B}}

\newcommand{\amdq}{\mathcal{A}^-\Delta q}
\newcommand{\apdq}{\mathcal{A}^+\Delta q}

\newcommand{\R}{\mathbb{R}}

\newcommand{\pmh}[1]{\ensuremath{#1 \frac{1}{2}}}
\newcommand{\xiph}{x_{\pmh{i+}}}
\newcommand{\ximh}{x_{\pmh{i-}}}
\newcommand{\xipmh}{x_{\pmh{i\pm}}}
\newcommand{\qiph}{q_{\pmh{i+}}}
\newcommand{\qimh}{q_{\pmh{i-}}}

\newcommand{\Dx}{\Delta x}
\newcommand{\chithree}{\chi^3}
\newcommand{\mx}{\ensuremath{m_x}}

\newcommand{\kphotonics}{Photonics Laboratory, King Abdullah University of Science and Technology \emph{(KAUST)}, Thuwal 21534, Saudi Arabia}
\newcommand{\numerics}{Numerical Mathematics Group, King Abdullah University of Science and Technology \emph{(KAUST)}, Thuwal 21534, Saudi Arabia}
\newcommand{\earc}{EXPEC Advanced Research Center, Aramco, Dhahran 31311, Saudi Arabia}

\graphicspath{{./figures/}}
\newcommand{\TheTitle}{A high-order finite volume method for Maxwell's equations in heterogeneous and time-varying media}
\newcommand{\TheAuthors}{D. P. San Roman Alerigi, D. I. Ketcheson, and B. S. Ooi}

\headers{
Maxwell's equations in heterogeneous and time-varying media}{D. P. San Roman Alerigi, D. I. Ketcheson, and B. S. Ooi}

\title{A high-order finite volume method for Maxwell's equations in heterogeneous and time-varying media}

\author{
    Damian P. San Roman Alerigi\thanks{
    \kphotonics~(\email{damian.sanroman@kaust.edu.sa}). Now at 
    \earc).
        }
    \and
    David I. Ketcheson\thanks{\numerics~(\email{david.ketcheson@kaust.edu.sa}).}
    \and
    Boon S. Ooi\thanks{\kphotonics~(\email{boon.ooi@kaust.edu.sa}).}
}

\ifpdf
\hypersetup{
  pdftitle={\TheTitle},
  pdfauthor={\TheAuthors}
}
\fi

\begin{document}
\maketitle
\begin{abstract}
We develop a finite volume method for Maxwell's equations in materials whose electromagnetic properties vary in space and time.  We investigate both conservative and non-conservative numerical formulations.  High-order methods are employed to accurately resolve fine structures that develop due to the varying material properties.  Numerical examples demonstrate the effectiveness of the proposed method in handling temporal variation and its efficiency relative to traditional 2nd-order FDTD.
\end{abstract}
%\tableofcontents
% REQUIRED

% REQUIRED

\section{Introduction\label{sec:introduction}}
Maxwell's equations in a charge- and current-free space are defined as:
\begin{subequations}\label{eq:introduction.maxwell}
    \begin{alignat}{1}
            \partial_t\emD - \nabla \times \emH &= 0 \label{eq:introduction.emE},\\
            \partial_t\emB + \nabla \times \emE &= 0 \label{eq:introduction.emH};
        \end{alignat}
\end{subequations}
where \ensuremath{\emE,\emB:\R^3\times\R^{3+1}\to\R^3} are the electric and magnetic fields, respectively. They are related to the material fields \ensuremath{\emD\in\R^3} and \ensuremath{\emH\in\R^3} via the constitutive relations:
\begin{subequations}\label{eq:introduction.constitutive}
        \begin{alignat}{1}
            \emD &= \emD \pare{\bare(\vec{r},t), \emE},\label{eq:introduction.cemD}\\
            \emB &= \emB \pare{\barm(\vec{r},t), \emH} \label{eq:introduction.cemB}.
        \end{alignat}
\end{subequations}
where \ensuremath{\bare,\barm:\R^{3\times 3}\times\R^{3+1}\to\R^{3\times 3}} are symmetrical and second-rank tensors with non-zero off-diagonal entries, and describe the electric and magnetic response of the material; they are known as permittivity and permeability.

In this work, we are interested in the general case in which \ensuremath{\bare,\barm} vary in space and time, and \emD, \emB~may be nonlinear.

We are motivated by recent interest in materials whose properties  can change at a pace comparable to that of electromagnetic wave oscillation; \eg \cite{Eggleton2010,Eggleton2011,San-Roman-Alerigi2013}. The effect of spatially heterogeneous materials on wave propagation has received much attention, but the effect of temporal material variations is much less studied. Novel effects  arising from wave propagation in space-time-varying media include trapping, confinement, ultra-short pulse coupling, beam transformation, negative refraction, eigen--polarization, and the optical Bohm--Aharonov effect \cite{Kaplan1983,Ridgely1999,Piwnicki2001,Rozanov2005,Leonhardt2006d,Leonhardt2006a,Rousseaux2008,Belgiorno2010,Cacciatori2010,Belgiorno2011}. This has opened the theoretical possibility to control light in new ways, \eg~ Alcubierre's warp drives \cite{Smolyaninov2011}.

A numerical scheme for electromagnetic wave propagation in time-dependent linear and heterogeneous media is developed in \cite{Farago2011}. The algorithm is based on the finite-difference time-domain (FDTD) method presented in \cite{KaneYee1966},  with an extension to account for the temporal variation of the medium. This is attained by applying an operator splitting method and Magnus' series expansion \cite{Farago2011}. Due to the splitting, the scheme requires that certain iteration matrices be recomputed at every time step.

We apply the high-order wave propagation method developed in \cite{Ketcheson2013}. This method is based on weighted essentially non-oscillatory (WENO) reconstruction, Runge-Kutta time integration, and wave-propagation Riemann solvers.  We expect these high-order algorithms to be more efficient than traditional second-order finite differences (i.e., the FDTD method) for high-frequency waves in rapidly-varying media. Furthermore, they can handle nonlinear media, do not require operator splitting,  and avoid the costly computation of iteration matrices required in \cite{Farago2011}.

The rest of the paper is organized as follows.  In Section \ref{sec:rewrite}, we consider two forms of Maxwell's equations with explicit time-dependence: first as a homogeneous and conservative hyperbolic system in which the conserved quantities appear implicitly, and second as a non-conservative system of balance laws. Section \ref{sec:waveprop} we review the high-order wave propagation method introduced in \cite{Ketcheson2013} and present the general algorithm to solve the Riemann problem. In Section \ref{sec:applications}, we derive the order of convergence for the algorithm using some fundamental 1D and 2D examples. Finally, in Section \ref{sec:summary}, we discuss future work, advantages, and opportunities.

\section{Maxwell's equations in nonlinear, time-varying media}\label{sec:rewrite}
To illustrate the difficulties involved in discretizing this system, let us consider the one-dimensional case.  Then \ensuremath{D, B, H, E} are scalar fields, and the system may be written in the form:
\begin{subequations}\label{eq:rewrite.em1D}
    \begin{alignat}{1}
        \partial_t{D} + \partial_x{H} &= 0, \label{eq:rewrite.em1D.E},\\
        \partial_t{B} + \partial_x{E} &= 0. \label{eq:rewrite.em1D.H},
    \end{alignat}
\end{subequations}

In general, the constitutive relations in \eqref{eq:introduction.constitutive} can take any shape. For the purpose of the following example, we follow a customary electromagnetic approach and expand them in a power series \cite{Griffiths1999}:
\begin{subequations}\label{eq:rewrite.constitutive-expand}
        \begin{alignat}{1}
            D \pare{\e (x,t), E} &= \eo\pare{E + \sum_i\chi_e^{(i)}(x,t)E^i}, \label{eq:rewrite.cemD-expand}\\
            B \pare{\mu(x,t), H} &= \mo\pare{H + \sum_i\chi_h^{(i)}(x,t)H^i}, \label{eq:rewrite.cemB-expand}.
        \end{alignat}
\end{subequations}
where \ensuremath{\eo,\mo\in\R} are the vacuum permittivity and permeability, and \ensuremath{\chi^{(n)}\in\R} are the electric (e) and magnetic (h) susceptibility scalar functions.

Equation \ref{eq:rewrite.constitutive-expand} allows us to  consider two types of constitutive relations. For linear media, we have
\begin{subequations}\label{eq:rewrite.linear-medium}
    \begin{align}
        D & = \e  E, \\
        B & = \mu H;
    \end{align}
\end{subequations}
whereas for non-linear meadia we can consider a system with cubic nonlinearity, \ie:
\begin{subequations} \label{eq:rewrite.nonlinear-medium}
    \begin{align}
        D & = \e  E + \chithree E^3, \\
        B & = \mu H + \chithree H^3.
    \end{align}
\end{subequations}

Finally, we observe that equation \ref{eq:rewrite.em1D} can be rewritten in conservative and non-conservative form.

\subsection{Conservative form}\label{sec:rewrite.conservative}
Let us define the conserved vector \q and flux \fq as:
\begin{align}\label{eq:rewrite.conservative.defs}
    \q  & = \begin{pmatrix} D \\ B \end{pmatrix}, &
    \fq & = \begin{pmatrix} H(\mu(x,t),B) \\ E(\e(x,t),D) \end{pmatrix};
\end{align}
hence we can summarize \eqref{eq:rewrite.em1D} as an homogeneous first-order hyperbolic system:
\begin{equation}\label{eq:rewrite.conservative.em1D}
    \partial_t \q + \partial_x \f(\q;x,t) = 0,
\end{equation}
with
\begin{align}\label{eq:rewrite.conservative.flux}
    \partial_q\f & = \begin{pmatrix} 0 & \partial_B H \\ \partial_D E & 0 \end{pmatrix},
\end{align}
and wave speeds (\ensuremath{s}) given by:
\begin{equation}\label{eq:rewrite.conservative.wave-speed}
    s = \pm \sqrt{\partial_D E \partial_B H }.
\end{equation}

\subsubsection{Linear media}\label{sec:rewrite.conservative.linear}
In the linear example we substitute \eqref{eq:rewrite.linear-medium} into \eqref{eq:rewrite.conservative.defs}, and obtain:
\begin{align}\label{eq:rewrite.conservative.linear.flux}
    \f(\q;x,t) & = \begin{pmatrix} B/\mu(x,t) \\ D/\e(x,t) \end{pmatrix}.
\end{align}

Consequently, the wave speeds are:
\begin{equation}\label{eq:rewrite.conservative.wspeed}
     s = \pm \sqrt{1/(\e \mu)}.
\end{equation}

\subsubsection{Nonlinear media}\label{sec:rewrite.conservative.nonlinear}
In the nonlinear case, we must determine derivatives of the inverse functions \ensuremath{H(\mu,B)} and \ensuremath{E(\e,D)} to compute the speed of the waves in \eqref{eq:rewrite.conservative.wave-speed} and other relevant quantities. We can approximate the terms \ensuremath{\partial_B H} and \ensuremath{\partial_D E} by means of an iterative method. For example, consider the nonlinear constitutive relation for the electric field given in  \eqref{eq:rewrite.nonlinear-medium}, and solve for \ensuremath{E(D)} as:
\begin{equation}\label{eq:rewrite.conservative.nonlinear.emEa}
    E(D) = \frac{1}{\e}\left(D - \chithree (E(D))^3\right).
\end{equation}
Substituting the latter into itself yields
\begin{equation}\label{eq:rewrite.conservative.nonlinear.emEb}
    E(D) = \frac{1}{\e}\left(D - \chithree D^3 + {\mathcal O}(D^5)\right).    
\end{equation}
Thus, finally we obtain the partial derivative of \ensuremath{E} with respect to \ensuremath{D} as:
\begin{equation}\label{eq:rewrite.conservative.nonlinear.emE_D}
    \partial_D E \approx \frac{1}{\e}\pare{1 - 3 \chithree_e D^2}.
\end{equation}

Similarly, we can follow the same algorithm to calculate the derivative of \ensuremath{H} with respect to \ensuremath{B} and obtain:
\begin{equation}\label{eq:rewrite.conservative.nonlinear.emH_B}
    \partial_B H \approx \frac{1}{\mu}\pare{1 - 3 \chithree_h B^2}.    
\end{equation}

So the wave speeds are:
\begin{equation}\label{eq:rewrite.conservative.nonlinear.wspeed}
    s \approx \pm\sqrt{\frac{1}{\e \mu}\pare{1 - 3 \chithree_e D^2}\pare{1 - 3 \chithree_h B^2}} + {\mathcal O}(D^3)
\end{equation}

\subsection{Non-conservative form}\label{sec:rewrite.nonconservative}
As we can see from the previous treatment, the conservative form can be intricate to solve complex nonlinear materials. An alternative is to derive a non-conservative form to \eqref{eq:rewrite.em1D}. In this case, we proceed by applying the chain rule to the time derivatives in \eqref{eq:rewrite.em1D}, obtaining:
\begin{subequations}\label{eq:rewrite.nonconservative.em1D}
    \begin{align}
        \partial_E D \partial_t E + \partial_x H &= -\partial_\e  D \partial_t \e, \\
        \partial_H B \partial_t H + \partial_x E &= -\partial_\mu B \partial_t \mu.
    \end{align}
\end{subequations}

Let us re-define the conserved quantity vector and flux as:
\begin{align}
    \q  & = \begin{pmatrix} E \\ H \end{pmatrix}, &
    \fq & = \begin{pmatrix} H \\ E \end{pmatrix};
\end{align}
and introduce a second-rank tensor known as the  capacity, \ensuremath{\bar{\kappa}}, and vector term named source, \ensuremath{\vec{\psi}} term:
\begin{align}
    \kap(\q;x,t)        &= \begin{pmatrix} \partial_E D \\ \partial_H B \end{pmatrix}, &
    \vec{\psi}(\q;x,t)  &= \begin{pmatrix} -E \partial_t \e \\ -H \partial_t \mu \end{pmatrix}.
\end{align}
Then, we can rewrite equation \ref{eq:rewrite.nonconservative.em1D} as a balance law of the form:
\begin{equation}\label{eq:rewrite.nonconservative.general}
    \kap(\q;x,t)\cdot\partial_t\q(x,t) + \partial_x\fq = \vec{\psi}(\q;x,t),
\end{equation}

This balance law results in a simpler form of linear and nonlinear cases. For linear media, substitute \eqref{eq:rewrite.linear-medium} into \eqref{eq:rewrite.nonconservative.em1D} to produce:
\begin{subequations}\label{eq:rewrite.nonconservative.linear}
    \begin{align}
        \e (x,t) E_t + \partial_x H &= -E \partial_t \e, \\
        \mu(x,t) H_t + \partial_x E &= -H \partial_t \mu.
    \end{align}
\end{subequations}
Whereas for nonlinear media we can substitute \eqref{eq:rewrite.nonlinear-medium} instead and obtain:
\begin{subequations}\label{eq:rewrite.nonconservative.nonlinear}
    \begin{align}
        \pare{\e  + 2 \chithree_e E^2} \partial_t E + \partial_x H &= -E\partial_t \e, \\
        \pare{\mu + 2 \chithree_h H^2} \partial_t H + \partial_x E &= -H\partial_t \mu.
    \end{align}
\end{subequations}

Note that in either case the speed of the waves is given
\begin{equation}\label{eq:rewrite.nonconservative.wspeed}
	s = \pm 1
\end{equation}

\section{Semi-discrete wave propagation} \label{sec:waveprop}
Numerical wave propagation methods for systems similar to \eqref{eq:rewrite.nonconservative.general} have been developed previously \cite{LeVeque2002,LeVeque2002book,Ketcheson2013}. However, in those methods, the capacity function $\kappa$ was assumed to depend only on $x$; whereas here, it may depend also on $q$ and $t$.  In this Section, we review those methods and discuss their extension to the case of \eqref{eq:rewrite.conservative.em1D} and \eqref{eq:rewrite.nonconservative.general}. Specifically, we extend the scheme presented in \cite{Ketcheson2013} to handle time-varying fluxes and capacity functions that depend on $q, x, t$.  Since that scheme is based on the method of lines, time-varying coefficients can be handled in a straightforward way. We focus first on the one-dimensional method and then briefly describe the extension to more spatial dimensions.

\subsection{Basic scheme}\label{sec:waveprop.scheme}
\begin{figure}[ht]
    \centering
    \begin{tikzpicture}
        \draw[step=2cm] (.1,-0.1) grid (5.9,2);
        \node at (2,-.25) {$x_{i-1/2}$};
        \node at (4,-.25) {$x_{i+1/2}$};
        \node at (3,1) {$i$};
        \node at (1,1) {$i-1$};
        \node at (5,1) {$i+1$};
    \end{tikzpicture}
    \caption{Schematic representation of 1D uniform cell grid.\label{fig:waveprop.grid1d}}
\end{figure}
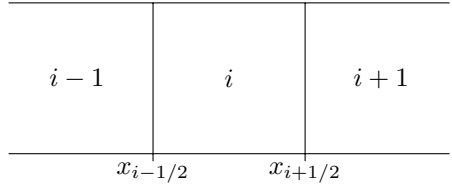

We consider a grid of cells with centers $x_i$ and interfaces $\xipmh$ (see figure \ref{fig:waveprop.grid1d}), and define the average of $q$ over cell $i$:
\begin{equation}
    Q_i = \int_{\ximh}^{\xiph} q(x,t) dx.
\end{equation}

From the cell averages we construct a piecewise-polynomial approximation to $q$:
\begin{equation}\label{eq:waveprop.scheme.polyq}
    \hat{\q}(x) = \hat{\q}_i \pare{x}, \qquad \text{for} \quad x\in\pare{\ximh,\xiph},
\end{equation}
where
\begin{equation}\label{eq:waveprop.scheme.hatq}
    \hat{\q}_i\pare{x} = \q\pare{x,t} + \mathcal{O}\pare{\Delta x^{r+1}}.
\end{equation}
At each interface $\ximh$, the approximation is discontinuous; hence, we define the interface values as:
\begin{align}\label{eq:waveprop.scheme.qlqr}
    q^L_\pmh{i-} & = \hat{q}_{i-1}(x_\pmh{i-}) & q^R_\pmh{i-} & = \hat{q}_i(x_\pmh{i-}).
\end{align}

\subsubsection{Conservative scheme}\label{sec:waveprop.scheme.conservative}
When applied to \eqref{eq:rewrite.conservative.em1D}, the semi-discrete scheme presented in \cite{Ketcheson2013} takes the form
\begin{equation}\label{eq:waveprop.scheme.conservative.qi}
    \frac{\partial Q_i}{\partial t} = - \frac{1}{\Dx} \left(\apdq_\pmh{i-} + \amdq_\pmh{i+} + f(q^L_\pmh{i+})-f(q^R_\pmh{i-})\right).
\end{equation}
The flux difference arises from integrating the flux term over the cell. The terms $\amdq, \apdq$ are referred to as fluctuations and incorporate the effects of waves emanating from the cell interfaces, where we must solve a Riemann problem. 

At each cell interface, we approximate the solution of a Riemann problem with initial states given by the reconstructed values at the interface:
\begin{equation}\label{eq:waveprop.scheme.conservative.initialconditions}
    q(x) = \left\{\begin{matrix} q^L_\pmh{i-} & x < \ximh, \\ q^R_\pmh{i-}  & x > \ximh. \end{matrix}\right.
\end{equation}
The solution is approximated by a sequence of jump discontinuities; this amounts to decomposing the jump as:
\begin{equation}\label{eq:waveprop.scheme.conservative.qlqr}
    q^R_\pmh{i-} - q^L_\pmh{i-} = \sum_{p=1}^m \alpha^p_\pmh{i-} r^p_\pmh{i-} = \sum_p \mathcal{W}^p_\pmh{i-}.
\end{equation}
Here the vectors $r^p$ are the eigenvalues of some approximate flux Jacobian at the interface. The vectors $\W^p$ are known as {\em waves}, and the collective effect of the right- and left-going waves yields the {\em fluctuations}:
\begin{subequations}\label{eq:waveprop.scheme.conservative.fluctuations}
    \begin{alignat}{1}
        \amdq_\pmh{i-} &\equiv \sum_p \pare{s^p_\pmh{i-}}^-\W^p_\pmh{i-},\label{eq:waveprop.scheme.conservative.fluctuations.left} \\
        \apdq_\pmh{i-} &\equiv \sum_p \pare{s^p_\pmh{i-}}^+\W^p_\pmh{i-},\label{eq:waveprop.scheme.conservative.fluctuations.right}
    \end{alignat}
\end{subequations} 
Here $s^p$ is the speed of the wave $\W^p$ (often given by an eigenvalue of an approximate flux Jacobian), and
\begin{equation*}
    \pare{s^p}^{-} = \min\pare{s^p,0} \qquad \pare{s^p}^{+} = \max\pare{s^p,0}.
\end{equation*}

\subsubsection{Non-conservative scheme}\label{sec:waveprop.scheme.nonconservative}
Following a similar derivation to its conservative counterpart, the non-conservative form \eqref{eq:rewrite.nonconservative.general} becomes 
\begin{equation}\label{eq:waveprop.scheme.nonconservative.qi}
    \frac{\partial Q_i}{\partial t} = -\frac{1}{\bar{K}_i\Delta x}\pare{\A^{+}\Delta q_{\pmh{i-}} + \A^{-}\Delta q_{\pmh{i+}} + f(q^L_\pmh{i+})-f(q^R_\pmh{i-})} + \Psi_i.
\end{equation}
Here $K_i$ and $\Psi_i$ are averages of $\kappa$ and $\psi$ over cell $i$.  For $\Psi$ we use the trapezoidal rule:
\begin{align}\label{eq:waveprop.scheme.nonconservative.psi}
    \Psi^1_{i} & = -\frac{\dot{\epsilon}^R_{\pmh{i-}}E^R_\pmh{i-}+\dot{\epsilon}^L_\pmh{i+} E^L_\pmh{i+}}{2} \\
    \Psi^2_{i} & = -\frac{\dot{\mu}^R_{\pmh{i-}}H^R_\pmh{i-}+\dot{\mu}^L_\pmh{i+} H^L_\pmh{i+}}{2}
\end{align}
where for readability $\dot\eta = \eta_t$.  For $\kappa$, we also use a trapezoidal average.
Let $\tilde{\eta}_i$ be a suitable average of  $\eta$ across the $i$th cell at some time $t^n$. Then, at the interface $x_{\pmh{i-}}$ we can approximate the average value $K_{\pmh{i-}}$ using the trapezoidal rule,
\begin{equation}\label{eq:waveprop.scheme.nonconservative.kappa}
    K_{k,\pmh{i-}}^n\approx\frac{\tilde{\eta}^R_{\pmh{i-}} + \tilde{\eta}^L_{\pmh{i-}}}{2} + \frac{p}{2^p}\:\chi^{(p)}\pare{q^R_{k,\pmh{i-}} + q^L_{k,\pmh{i-}}}^{p-1}, \quad p\geq 2,
\end{equation}

\subsection{Riemann solver}\label{sec:waveprop.riemann}
Following (LeVeque, 2002), we can design a Riemann solver for this system as follows. We assume the values of $\epsilon, \mu$ are constant in each cell so that we can write $\epsilon_i, \mu_i$. The eigenvector matrix for $f_q$ is
\begin{align}\label{eq:waveprop.riemann}
    R_\pmh{i-} & = \begin{pmatrix} -Z_{i-1} & Z_i \\ 1 & 1 \end{pmatrix},
\end{align}
where $Z_i = \sqrt{H_B/E_D}$.  We can use an $f$-wave solver by solving the system
\begin{equation}\label{eq:waveprop.riemann.beta}
    R\beta = f_i(Q_i) - f_{i-1}(Q_{i-1}).
\end{equation}

%(need to expand this section a bit)

\subsection{Reconstruction and time integration}\label{sec:waveprop.reconstruction}
It remains to specify the method of time integration and spatial reconstruction. These ingredients determine the order of accuracy of the scheme.

We use the ten-stage fourth-order strong-stability-preserving Runge-Kutta scheme described in \cite{Ketcheson2008} for time integration. Although it uses many stages, this method has a large region of absolute stability, allowing the use of large CFL numbers.

We use fifth-order component-wise WENO reconstruction in space as described in \cite{Shu2009}. Traditionally, methods based on finite differences use several points per wavelength  to correctly model the high-frequency waves that occur in some physical problems, \eg electromagnetism. We use  WENO or any high-degree polynomial  because it allows us to keep a high level of accuracy with relatively coarser grids. 

All of this is implemented in Clawpack's PyClaw module \cite{Ketcheson2012,clawpack2014}. The Riemann solvers for Maxwell's equations are implemented in a separate package called EMClaw, available at \cite{emclaw2016}.

\subsection{Algorithm}\label{sec:waveprop.algorithm}
At every Runge-Kutta stage, the numerical implementation of \eqref{eq:waveprop.scheme.nonconservative.qi} follows the steps:
\begin{enumerate}
    \item For the non-conservative system only: update cell averages of the time-dependent capacity $\bar{K}_i^n$, and source $\Psi_i^n$ functions;
    \item Reconstruct the interface values $\qimh^R,\qiph^L$ using fifth-order WENO reconstruction;
    \item Solve the Riemann problem with initial states $(\qimh^L,\qimh^R)$ to compute the fluctuations, $\A^{\pm}\Delta\qimh$;
    \item Calculate the flux difference $f(q^L_\pmh{i+})-f(q^R_\pmh{i-})$;
    \item Compute $\partial Q_i /\partial t$ using the semi-discrete scheme \eqref{eq:waveprop.scheme.nonconservative.qi}.
\end{enumerate}

\subsection{Extension to two dimensions}\label{sec:waveprop.extendnD}
Using a dimension-by-dimension approach, we extend the numerical wave propagation method to two dimensions where the main equation is (on the Cartesian grid):
\begin{equation}\label{waveprop.extend.general}
    \kap\cdot\q_t + \vec{f}(\q)_x + \vec{g}(\q)_y = \vec{\psi}(\q,x,y,t).
\end{equation}

In two dimensions, the semi-discrete scheme \eqref{eq:waveprop.scheme.nonconservative.qi} takes the form:
\begin{equation}\label{eq:waveprop.extendnD.2d}
    \begin{split}
    \frac{\partial Q_{ij}}{\partial t} = -\frac{1}{\bar{K}_{ij}\Delta x \Delta y}
        \Bigg(&\A^{+}\Delta q_{\pmh{i-},j} + \A^{-}\Delta q_{\pmh{i+},j} +
        f(q^L_{\pmh{i+},j}) - f(q^R_{\pmh{i-},j})  \\
            &+\B^{+}\Delta q_{i,\pmh{j-}} + \B^{-}\Delta q_{i,\pmh{j+}} +
        g(q^L_{i,\pmh{j+}}) - g(q^R_{i,\pmh{j-}})\Bigg)  \\
            & + \Psi_{ij}.
    \end{split}
\end{equation}

The fluctuation terms are determined by solving the Riemann problem in the corresponding direction and initial data; for example, $\B^{+}\Delta q_{i,\pmh{j-}}$ is calculated by solving the Riemann problem in the $y$-direction with initial states $(q^L_{i,\pmh{j-}},q^R_{i,\pmh{j-}})$. As noted in \cite{Ketcheson2013}, for the method to be high-order accurate, the fluctuations and flux differences must be computed based on a multidimensional reconstruction of $q$.  In the present work, we instead use a much cheaper dimension-by-dimension reconstruction, described in \cite{Ketcheson2013}. Although this leads to a formally 2nd-order accurate scheme, it is still much less dissipative than traditional second-order schemes and allows high-frequency waves to be resolved on much coarser grids.

\section{Numerical applications and convergence}\label{sec:applications}
We test the above-described techniques numerically by studying the propagation of electromagnetic waves in media with spacetime variations. We choose examples where the exact solution can be  calculated with great accuracy.

To obtain the results in this section, we developed EMClaw \cite{emclaw2016},  a multi-dimensional numerical solver for Maxwell's equations in spacetime-varying and nonlinear media, based on the scheme \eqref{eq:waveprop.scheme.nonconservative.qi}. EMClaw was designed as an extension of the Clawpack package \cite{clawpack2014}.

\subsection*{Miscellaneous definitions}\label{sec:applications.definitions}
We use the numerical examples to measure the $L_1$ errors, \emE, and rate of convergence, $p$, of the numerical schemes developed in previous sections. Specifically, we compute
\begin{equation}
    E_{L_1} = \Delta x \sum_i|Q_i - \hat{Q_i}|,
\end{equation}
where $\hat{Q_i}$ is a highly accurate solution cell average computed by characteristics (exact solution) or by using grid refinement.

For the subsequent numerical studies, let $E_e(h)$  denote the error of the numerical solution with respect to the exact solution for some grid spacing $h$, and similarly, let us denote by $E_s(h)$ the error of the numerical solution on a grid with spacing $h$ with respect to the numerical solution obtained on the refined grid, $h/2$. 

Let $I$ denote the 2-norm of $\q$, the magnitude of the field,
\begin{equation}\label{eq:applications.definitions.intensity}
    I(\vec{r},t) = \sqrt{\sum_{i=1}^m q_i^2(\vec{r},t)}.
\end{equation}
Similarly, let $n$ denote the magnitude of the material profile 
\begin{equation}\label{eq:applications.definitions.n}
    n=\sqrt{\eta_e\cdot\eta_h}.
\end{equation} 
Finally let $F=I,n$, then we can define the maximum $F_{max}(t)=\max_{\vec{r}}{F(\vec{r},t)}$,  and the path described by it $\vec{r}_F(t)=\{\vec{r}:F(\vec{r},t)=F_{max}\}$.

\subsection{1D Electromagnetic}\label{sec:applications.1D}
Here we apply scheme \eqref{eq:waveprop.scheme.nonconservative.qi} to one-dimensional Maxwell's equations in the case of spacetime-varying media with linear and nonlinear components. We assume the materials to be isotropic and piecewise homogeneous, leading to the 1D system \eqref{eq:rewrite.nonconservative.em1D}.

For the examples below, we set the initial condition to be the right-moving pulse
\begin{equation}\label{eq:applications.1D.qo}
    E(x,0)= H(x,0) = \exp{\pare{\frac{x-x_o}{\sigma}}^2}.
\end{equation}
Reflecting boundary conditions are used in all 1D examples.
 
\subsubsection{Time-varying medium}\label{sec:applications.1D.sin}
We consider 1D electromagnetic waves in a  linear, spatially homogeneous, and time-varying medium. Namely, we solve \eqref{eq:rewrite.nonconservative.em1D} in the interval $x\in[0,100]$, with
\begin{equation*}
    \e = \mu = 1 + \delta\eta\sin\pare{\frac{12\pi}{100} t},
\end{equation*}
where $\delta\eta\in\R$ is the oscillation amplitude.

The initial condition is given by \eqref{eq:applications.1D.qo} with $x_o=10$ and $\sigma=2$.

% (introduce exact solution)

\begin{figure}[h!]
    \begin{center}
    \includegraphics[width=0.5\textwidth]{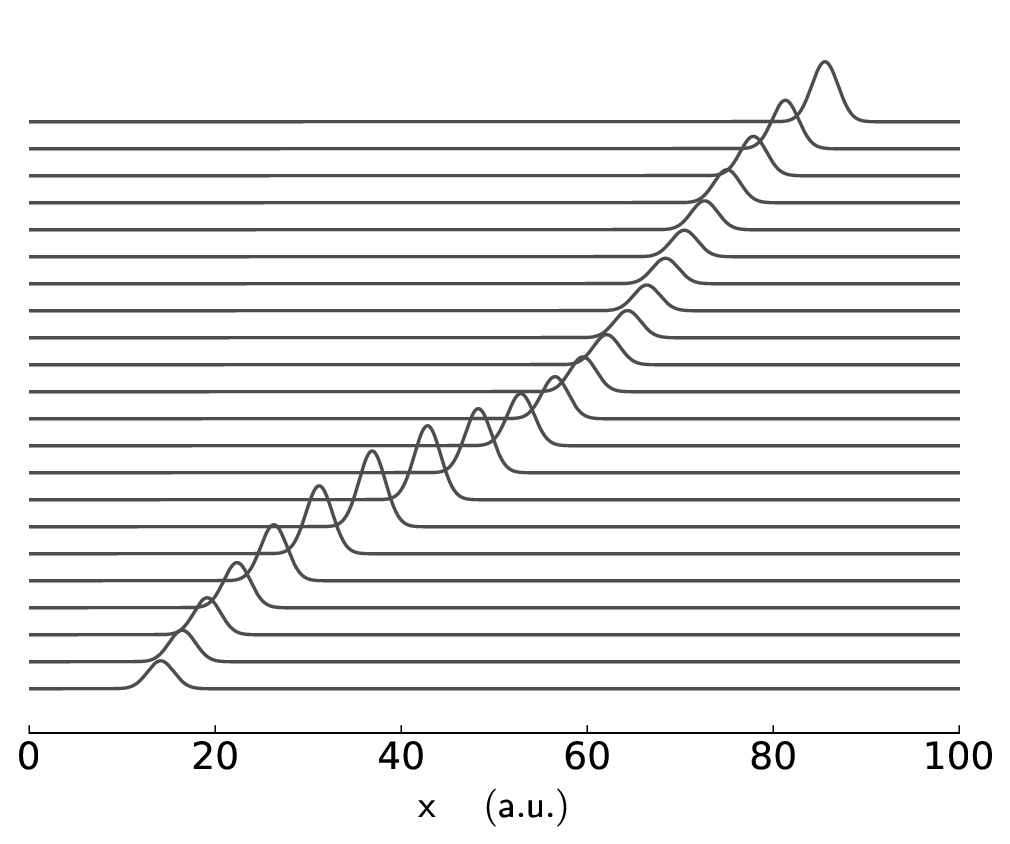}
    \caption{Time-staggered plot of $I$ for 1D wave propagation in a vibrating medium with $n_{\omega}=12$.\label{fig:applications.1D.sin.waterfall}}
    \end{center}
\end{figure}

Table \ref{table:applications.1D.sin.errors} shows the error and convergence rates for propagation in this problem at different grid sizes $h$ defined by the number of cell grids \mx.

\begin{table}[h!]
    \centering
    \begin{tabular}{r|r|rr|rr}
        \hline
        \multicolumn{1}{c}{~} &   \multicolumn{1}{c}{~}  & \multicolumn{2}{c}{Conservative} & \multicolumn{2}{c}{Non-Conservative} \\
        \multicolumn{1}{c}{$m_x$} &   \multicolumn{1}{c}{$h$} &   \multicolumn{1}{c}{$E_e(h)$} & \multicolumn{1}{c}{$p_e$} &  \multicolumn{1}{c}{$E_e(h)$} &  \multicolumn{1}{c}{$p_e$} \\
        \hline
        128     & 7.812e-01 &  2.814e-01 & 3.657 & 2.821e-01 & 3.597 \\
        256     & 3.906e-01 &  2.230e-02 & 3.904 & 2.331e-02 & 5.519 \\
        512     & 1.953e-01 &  1.490e-03 & 2.457 & 5.085e-04 & 5.713 \\
        1024    & 9.766e-02 &  2.714e-04 & 5.819 & 9.694e-06 & 5.265 \\
        2048    & 4.883e-02 &  4.809e-06 & 3.820 & 2.521e-07 & 5.078 \\
        4096    & 2.441e-02 &  3.405e-07 & 6.160 & 7.464e-09 & 5.028 \\
        \hline
    \end{tabular}
\caption{Errors and convergence rate for 1D electromagnetic wave propagation in a linear, spatially homogeneous, and time-varying medium. Comparison between the Conservative and Non-Conservative Riemann solvers, with $cfl=2.4$.\label{table:applications.1D.sin.errors}}
\end{table}

\begin{figure}[h!]
    \centering
    \subfloat[Conservative]{\includegraphics[width=0.5\textwidth]{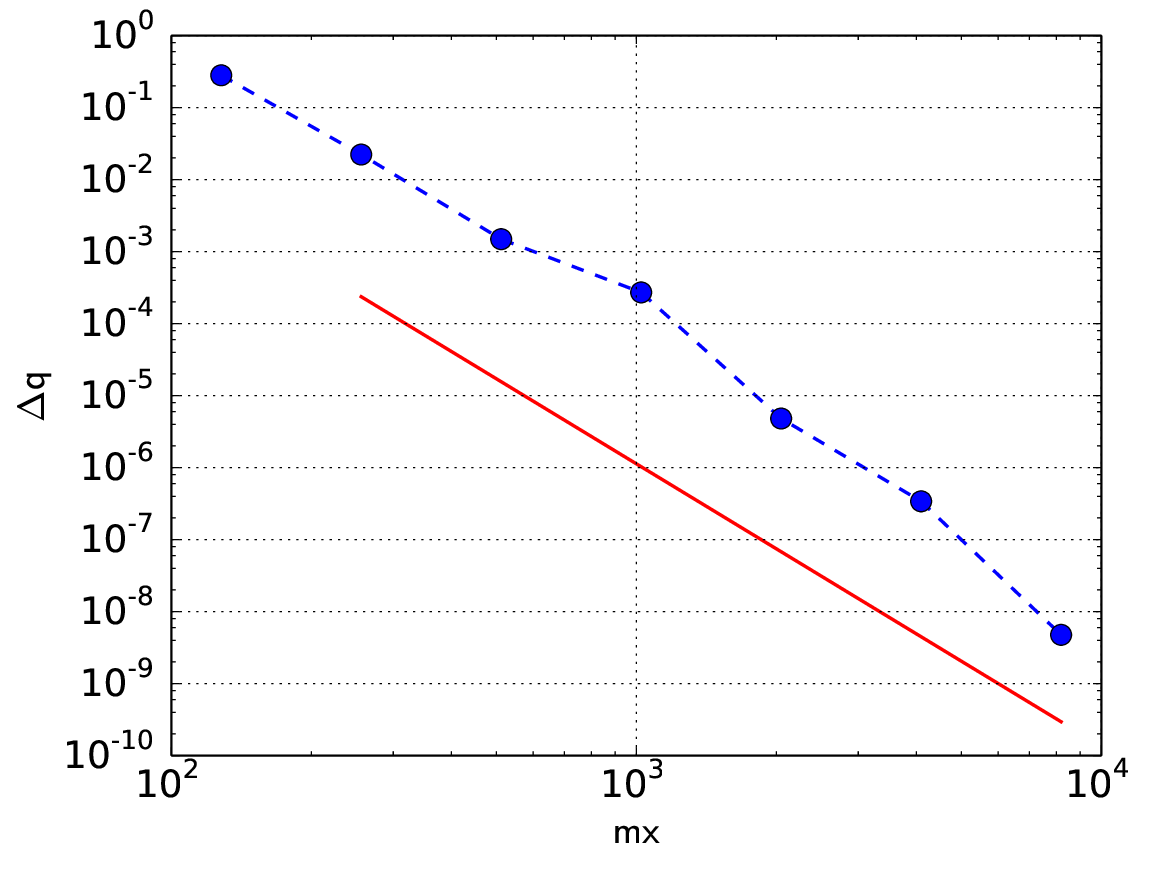}}
    \subfloat[Non-Conservative]{\includegraphics[width=0.5\textwidth]{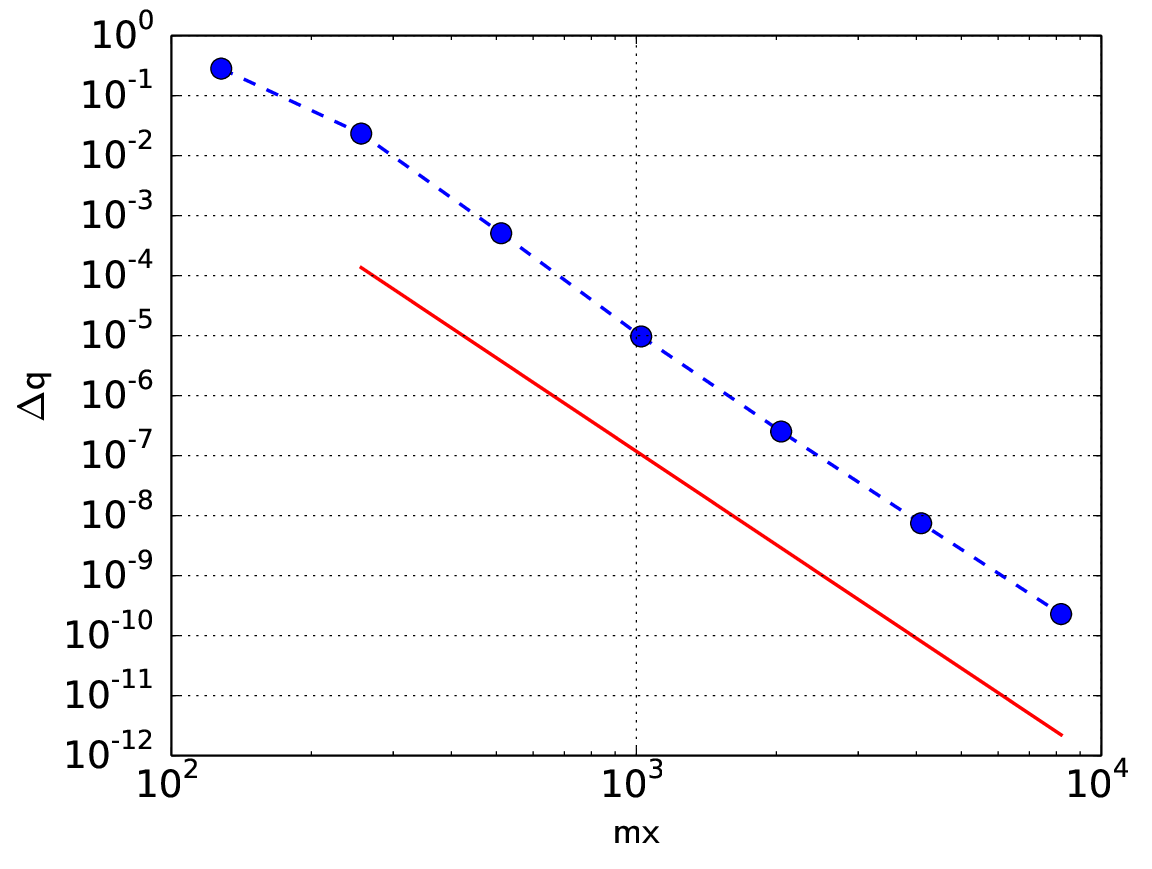}}
    \caption{Convergence of the finite volume solution to the analytic solution, $E_e(h)$,  for 1D electromagnetic wave propagation in a linear, spatially homogeneous and time-varying medium (blue-dashed); comparison between the conservative and non-conservative Riemann solvers with $cfl=2.4$. Linear fit with slope (red) $p_e\approx -5.173$ (non-conservative) $p_e \approx -3.922$ (conservative).\label{fig:applications.1D.sin.convergence}}
\end{figure}

Figure \ref{fig:applications.1D.sin.convergence} shows $E_e(h)$ for this case as a function of $\mx = L/h$, where $L=100$ is the length of the simulation space. Using a least-square method, we can approximate the slope of the line in the asymptotic convergence region of $E_e(h)$ and  find $p_e \approx -3.922$ for the conservative, and $p_e\approx 5.173$ for the non-conservative, Riemann solvers.

\subsubsection{Spacetime-varying medium}\label{sec:applications.1D.spt}
We now consider 1D electromagnetic waves in a  linear and spacetime-varying medium. Namely, we solve 1D equation \eqref{eq:rewrite.em1D} in the interval $x\in[0,300]$, with
\begin{equation}\label{eq:applications.1D.spt.rip}
    \eta = \eta_o + \delta\eta\exp{\pare{\frac{x-x_o-v\:t}{\sigma}}^2},
\end{equation}
where where $\eta_o\in\R$ is the background material parameter, $\delta\eta\in\R$ is the spacetime-variation's amplitude, $x_o\in\R$ is the offset, $v\in\R$ is the velocity, and $\sigma\in\R$ is the full-width half maximum. 

To be specific, we set $\eta_o=1.5$, $\delta\eta=0.15$, $x_o=25$, $v=0.59$, and $\sigma=5$; for the initial conditions set $q_o$ with $x_o=10$ and $\sigma=2$.

\begin{figure}[h!]
    \centering
    \includegraphics[width=0.5\textwidth]{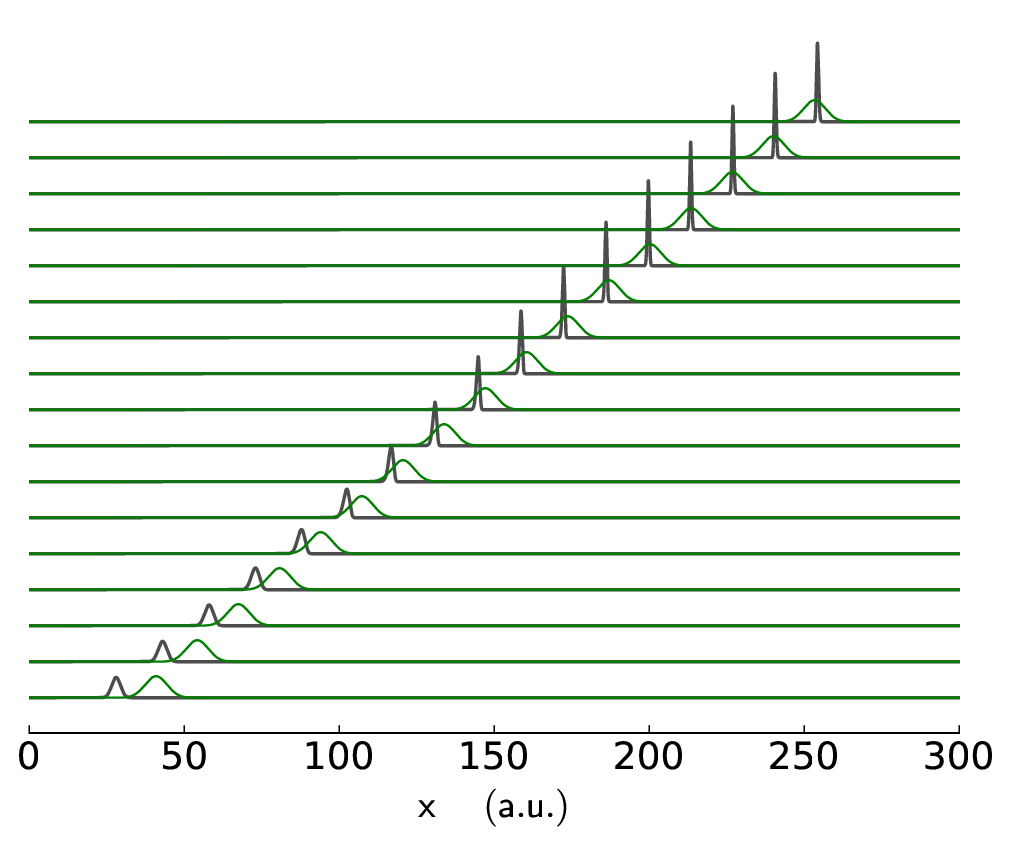}
    \caption{Time-staggered plots of  $I$ (black) and $n$ (green) for 1D wave propagation in a Gaussian-like spacetime-varying medium.\label{fig:applications.1D.spt.waterfall}}
\end{figure}

Table \ref{table:applications.1D.spt.errors} shows the errors and convergence rates for propagation in this problem.

\begin{table}[h!]
    \centering
    \begin{tabular}{r|r|rr|rr}
        \hline
        \multicolumn{1}{c}{~} &   \multicolumn{1}{c}{~}  & \multicolumn{2}{c}{Conservative} & \multicolumn{2}{c}{Non-Conservative} \\
        \multicolumn{1}{c}{$m_x$} &   \multicolumn{1}{c}{$h$} &   \multicolumn{1}{c}{$E_e(h)$} & \multicolumn{1}{c}{$p_e$} &  \multicolumn{1}{c}{$E_e(h)$} &  \multicolumn{1}{c}{$p_e$} \\
        \hline
        128     & 7.812e-01 &   3.648e+00 & 1.369 & 3.775e+00 & 1.365 \\
        256     & 3.906e-01 &   1.413e+00 & 1.411 & 1.465e+00 & 1.407 \\
        512     & 1.953e-01 &   5.315e-01 & 1.575 & 5.524e-01 & 1.544 \\
        1024    & 9.766e-02 &  1.784e-01  & 2.677 & 1.895e-01 & 2.658 \\
        2048    & 4.883e-02 &  2.790e-02  & 2.559 & 3.002e-02 & 4.185 \\
        4096    & 2.441e-02 &  4.735e-03  & 1.941 & 1.650e-03 & 5.217 \\
        \hline
    \end{tabular}
    \caption{Errors and convergence rate for 1D wave propagation in a Gaussian-like spacetime-varying medium. Comparison between the Conservative and Non-Conservative Riemann solvers, with $cfl=2.4$\label{table:applications.1D.spt.errors}}
\end{table}

\begin{figure}[h!]
    \centering
    \subfloat[Conservative]{\includegraphics[width=0.5\textwidth]{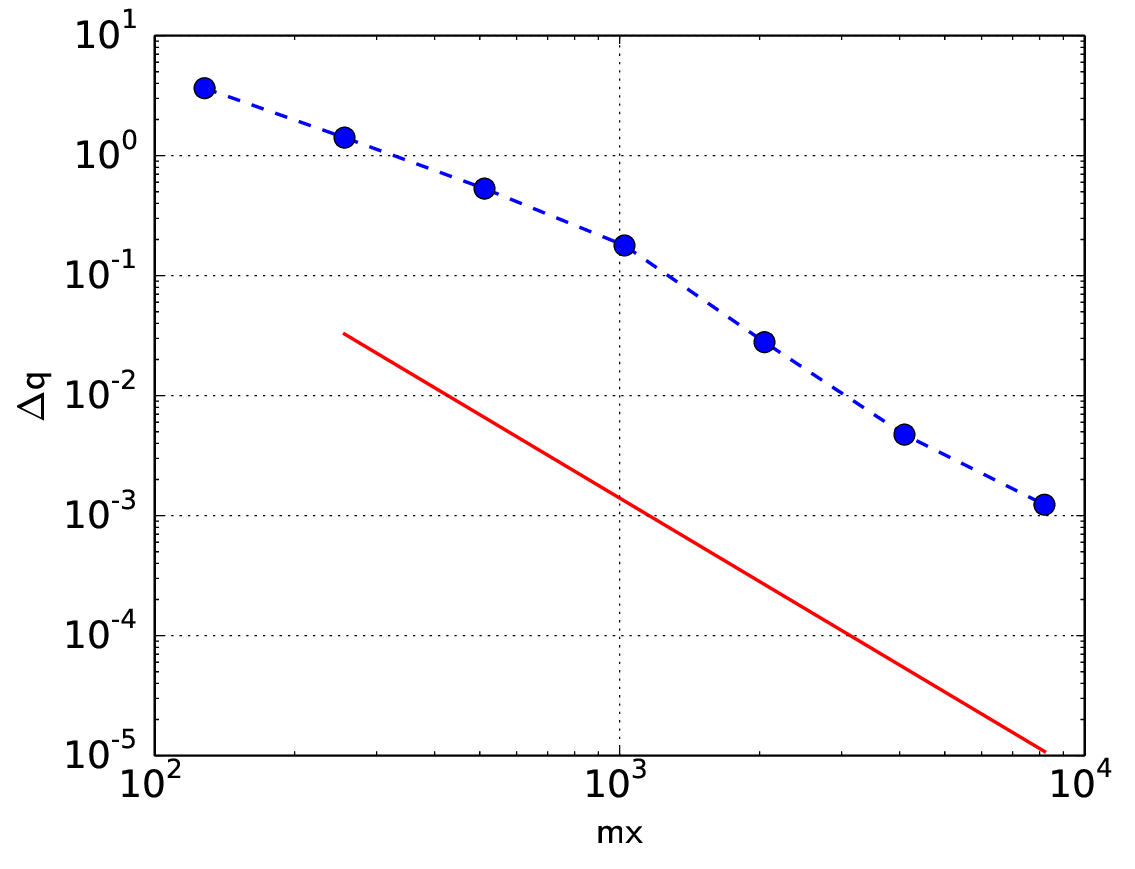}}
    \subfloat[Non-Conservative]{\includegraphics[width=0.5\textwidth]{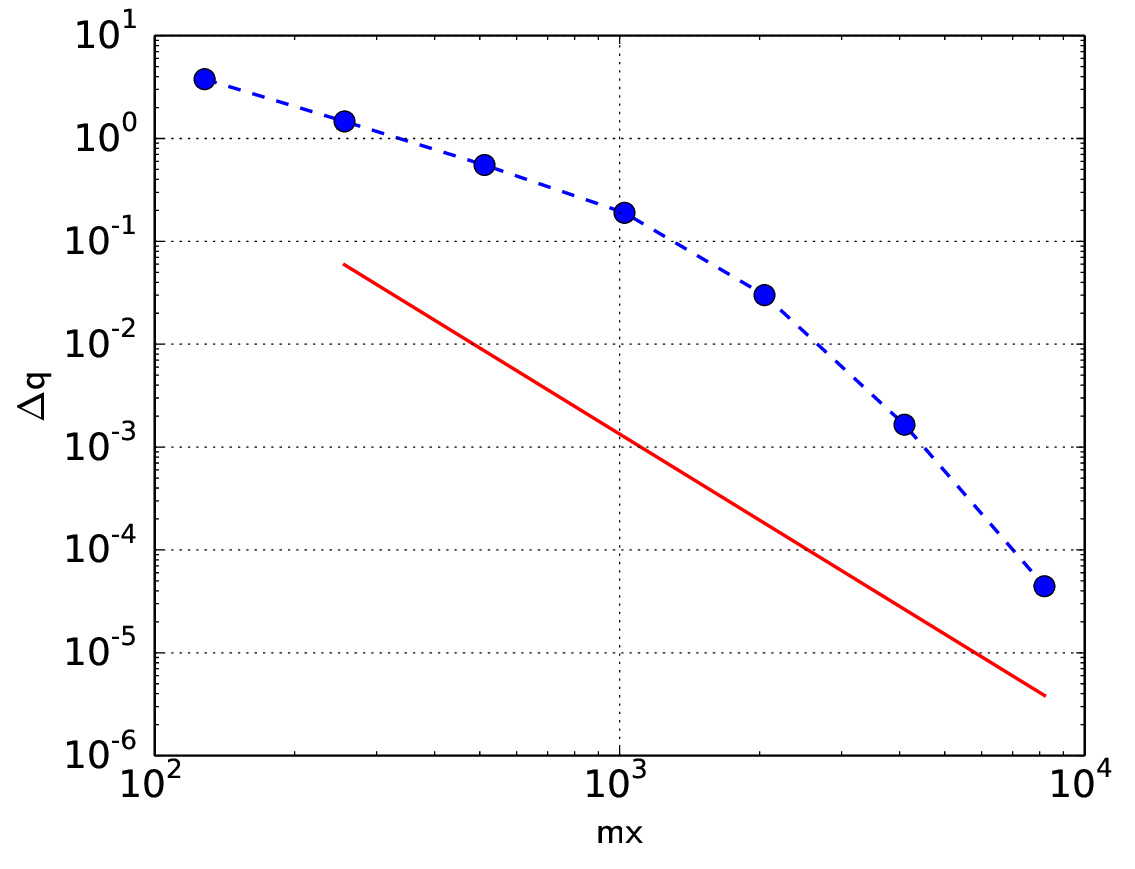}}
    \caption{Convergence of the finite volume solution to the analytic solution, $E_e(h)$,  for 1D electromagnetic wave propagation in a linear, spatially homogeneous and time-varying medium (blue-dashed); comparison between the conservative and non-conservative Riemann solvers with $cfl=2.4$. Linear fit with slope (red) $p_e\approx -2.782$ (non-conservative) $p_e \approx -2.311$ (conservative).\label{fig:applications.1D.spt.convergence}}
\end{figure}

Figure \ref{fig:applications.1D.spt.convergence} shows $E_e(h)$ for  1D wave propagation in a Gaussian-like spacetime-varying medium as a function of $\mx = L/h$, where $L=300$ is the length of the simulation space. Using a least-square method, we can approximate the slope of the line in the asymptotic convergence region of $E_e(h)$ and  find $p_e\approx 3.481$. 

To explain the difference in the rate of convergence between the 1D wave propagation problems,  recall that in the former, the material is only time-dependent, which together with the results would suggest that the discrepancy is due to the second-order approximation to $\Psi_i$  and $K_i$ within each cell $i$ for spacetime-varying media. Note that because we use the trapezoidal rule,  we would expect the convergence rate to be second-order accurate; the higher degree of convergence ensues from using high-order WENO interpolation and Runge-Kutta methods.

\subsubsection{Nonlinear flowing medium}\label{sec:applications.1D.nonlinearspt}
In this section, we study a hybrid material that incorporates the spacetime-varying medium of the previous section and a background nonlinear material, $\chi^{(3)}\neq 0$. This problem is difficult to study in the paraxial approximation, and a numerical scheme is usually required.  

The material parameters and domain are the same in this example as in the previous section. The nonlinearity is introduced by setting  $\bar{\chi}^{(3)}=0.1$. The initial condition $q_o$ is again \eqref{eq:applications.1D.qo} with $x_o=10$ and $\sigma=2$.

\begin{figure}[h!]
    \centering
    \subfloat[Nonlinear and stationary medium]{\includegraphics[width=0.4\textwidth]{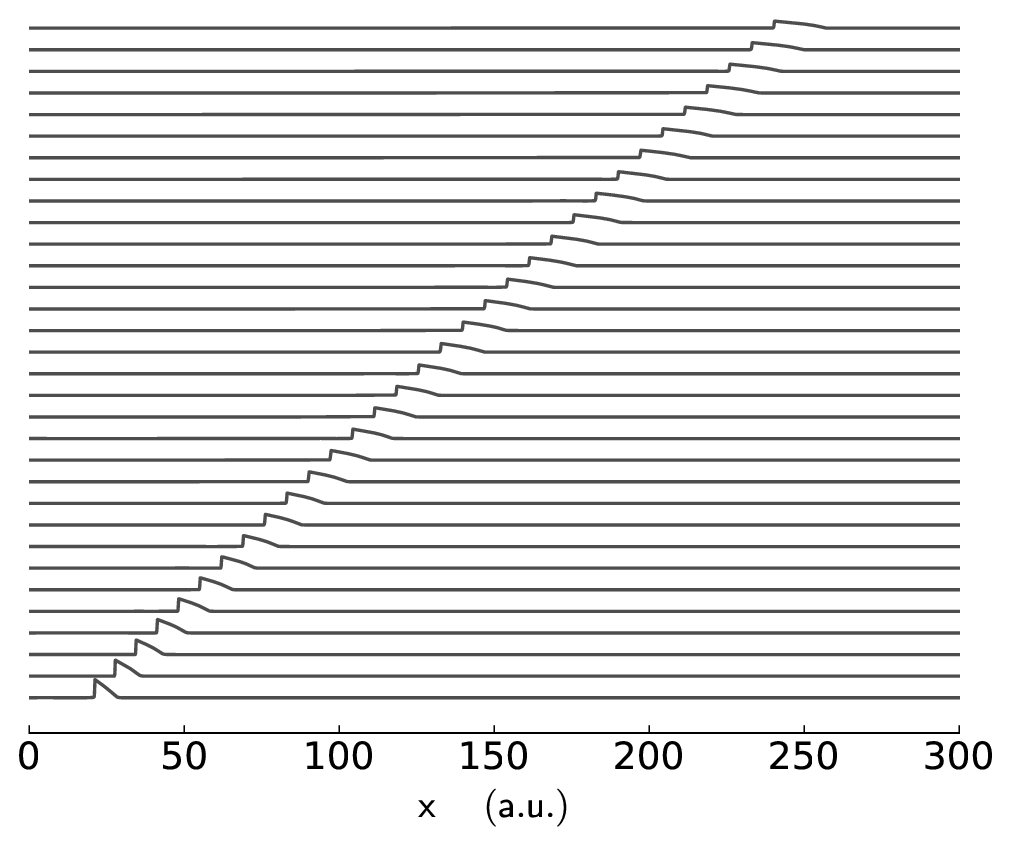}}
    \subfloat[Nonlinear and spacetime-varying medium]{\includegraphics[width=0.4\textwidth]{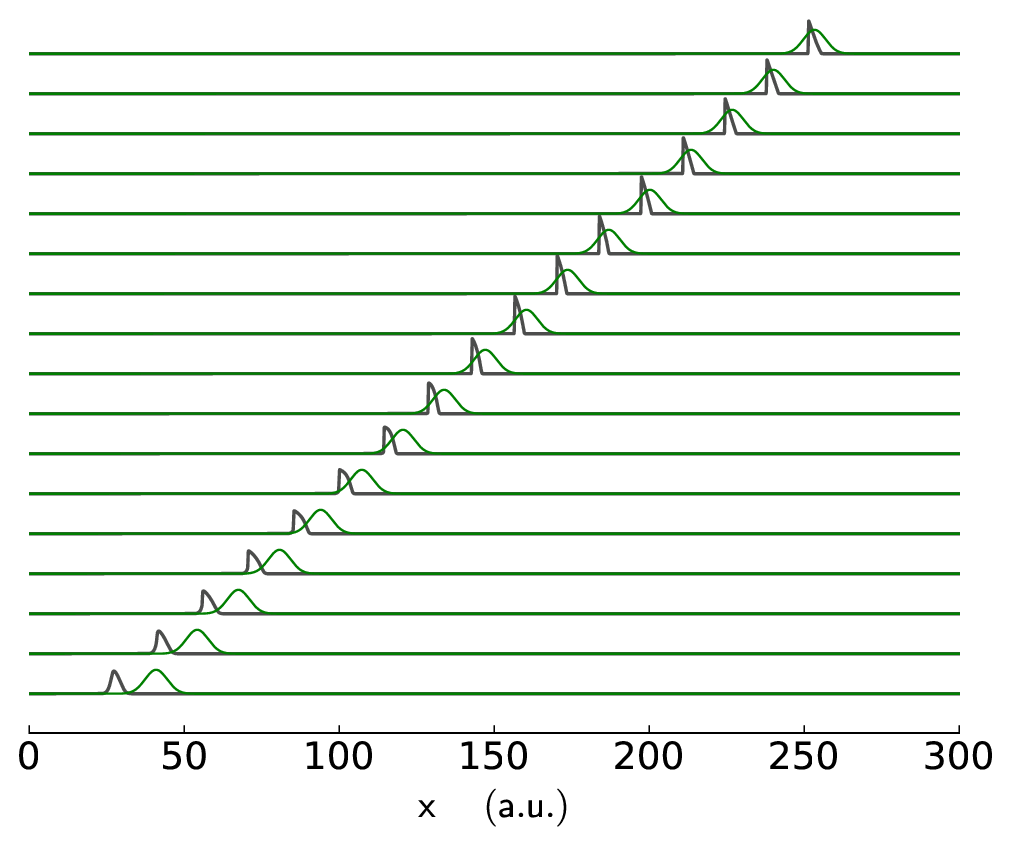}}
    \caption{Time staggered  plot of $I$ for 1D wave propagation in a nonlinear and homogeneous medium (left), and a nonlinear spacetime-varying medium (right).\label{fig:applications.1D.nonlinearspt.waterfall}}
\end{figure}

Figure \ref{fig:applications.1D.nonlinearspt.waterfall} (a) shows the effect of the nonlinearity in the absence of the moving perturbation \eqref{eq:applications.1D.spt.rip}. Note that as the pulse evolves, the nonlinearity in the medium compresses the pulse where $q_x \geq 0$ and expands it where $q_x\leq 0$; that is, we observe the formation of a shock and rarefaction, respectively.

Allow us now to introduce the flowing material of equation \ref{eq:applications.1D.spt.rip} and observe the evolution of the right-moving pulse \eqref{eq:applications.1D.qo} plotted in Figure \ref{fig:applications.1D.nonlinearspt.waterfall} (b). The results  suggest that the moving perturbation balances the dispersion introduced by the nonlinearity; in other words,  the former causes the front of the pulse to expand while the moving perturbation compresses it.

\begin{figure}[h!]
    \centering
    \subfloat[{$\chi^{(3)}=0$}]{\includegraphics[width=0.45\textwidth]{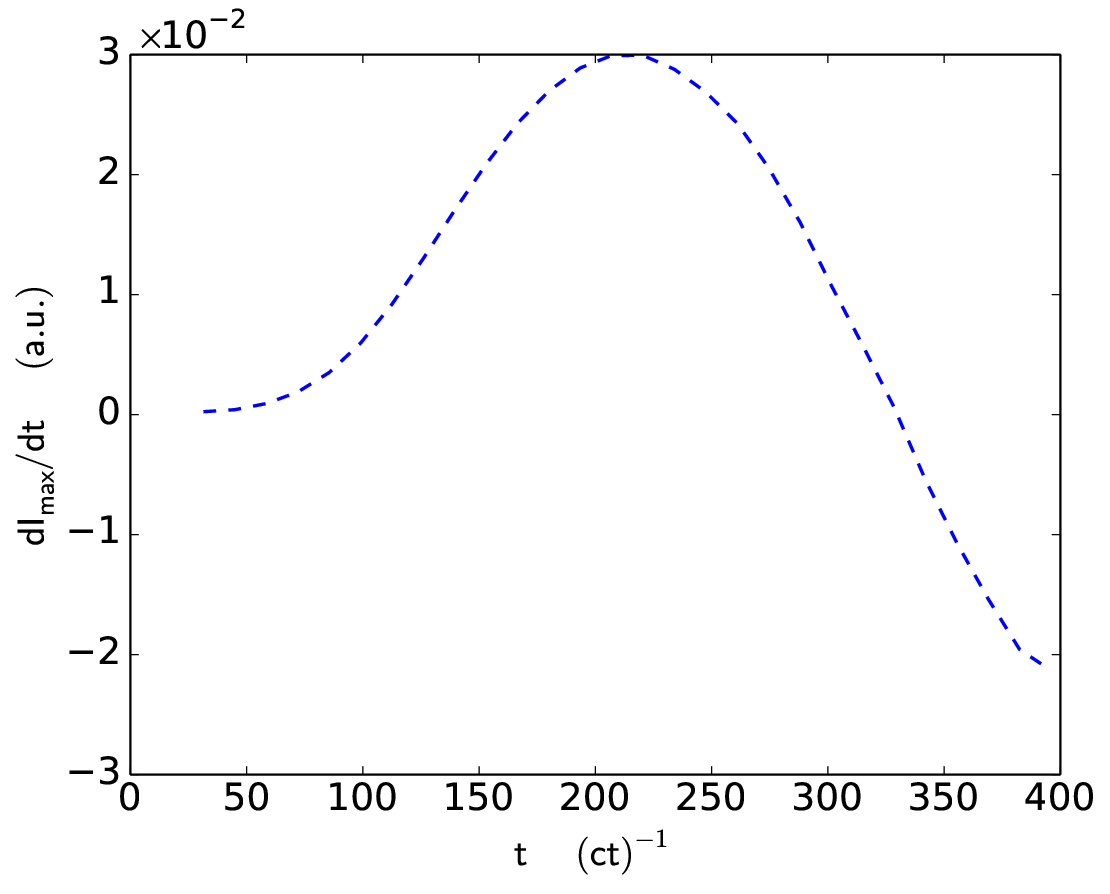}}
    \subfloat[{$\chi^{(3)}=0.1$}]{\includegraphics[width=0.45\textwidth]{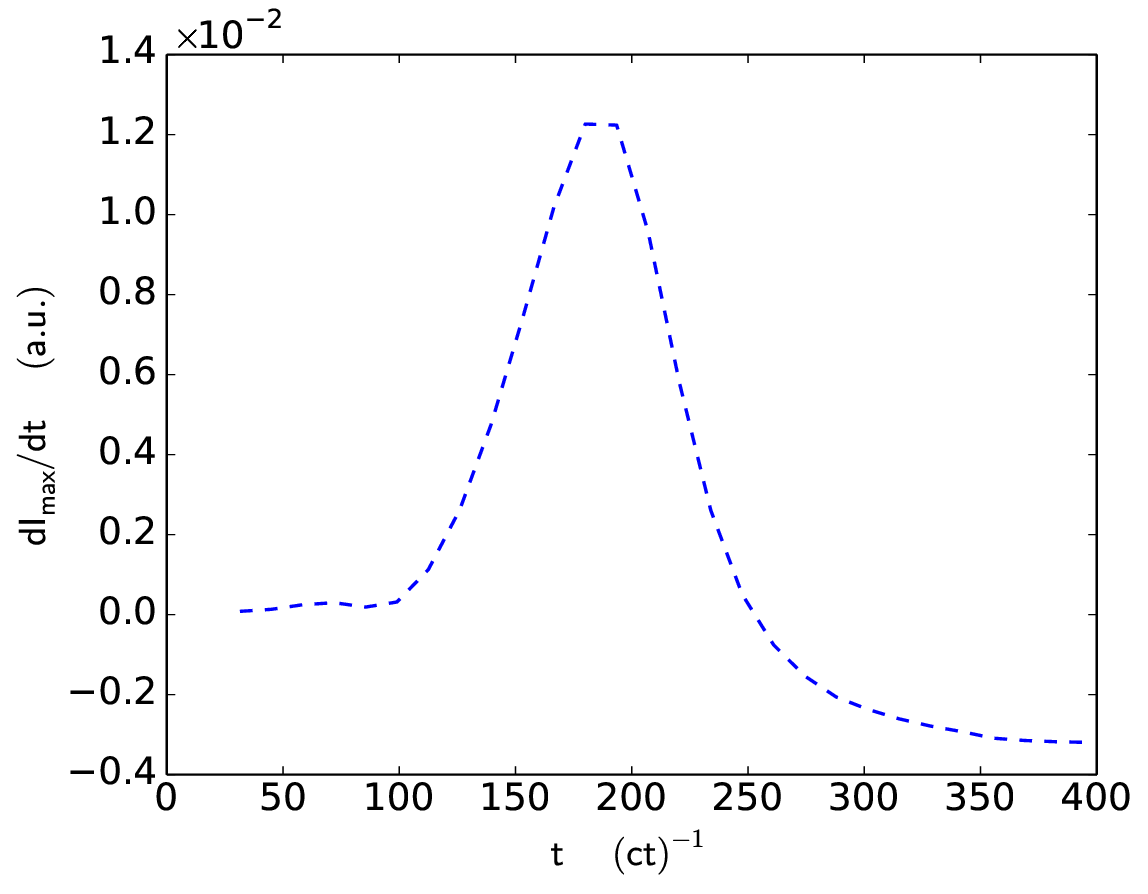}}
    \caption{Time rate of change of $I_{max}$ for 1D wave propagation in a nonlinear Gaussian-like spacetime-varying medium.\label{fig:applications.1D.nonlinearspt.didt}}
\end{figure}

To better appreciate this dynamic, let us look at the time rate of change of $I_{max}$ plotted in Figure \ref{fig:applications.1D.nonlinearspt.didt}, notice that as the pulse interacts with the perturbation, it begins to compress and increase in amplitude, while at the same time, the nonlinearity causes rarefaction; ultimately, this leads to equilibrium, as exemplified by the fact that $\frac{dI_{max}}{dt}\approx 0$.
  
\subsection{2D Electromagnetic}\label{sec:applications.2D}
Here we apply scheme \eqref{eq:waveprop.extendnD.2d} to the two-dimensional electromagnetic problem in the case of linear spacetime-varying media. We assume the materials to be isotropic and piecewise homogeneous, leading to the 2D system:
\begin{subequations}\label{eq:applications.2D.general}
    \begin{alignat}{1}
        \kappa_1(q_1,x,t)\pare{q_1}_t -  \pare{q_3}_y&= \psi_e(q_1,x,t), \\
        \kappa_2(q_2,x,t)\pare{q_2}_t + \pare{q_3}_x&= \psi_e(q_2,x,t), \\
        \kappa_3(q_3,x,t)\pare{q_3}_t + \pare{q_2}_x - \pare{q_1}_y&= \psi_h(q_3,x,t);  
    \end{alignat}
\end{subequations}
with initial condition to be the right-moving pulse
\begin{subequations}\label{eq:applications.2D.qo}
    \begin{alignat}{1}
        q_0(x,0) &= 0.0,\\
        q_1(x,0) &= g(y,y_o)\:q_o(x,x_o,\sigma),\\
        q_2(x,0) &= g(y,y_o)\:q_o(x,x_o,\sigma),
    \end{alignat}
\end{subequations}
where $\q_o$ is defined following \eqref{eq:applications.1D.qo}, and $g:\R\to\R$ defines the transversal profile of the pulse. To be specific, we choose
\begin{equation*}
    g(y) = \cos{\frac{\pare{y-y_o}\pi}{L_y}}
\end{equation*}
where $L_y$ is the simulation length in the $y$ direction.

We solve \eqref{eq:applications.2D.general} in the interval $\{x,y\}\in[0,180]\times[0,180]$ with varying grid space and set the all the boundary conditions to be reflecting or wall.

For the sake of simplicity, we again consider propagation in a system where $\eta_l^0=1$ and the material has unitary impedance, \ie $\eta_k = \eta$. Namely, the material profile is 
\begin{equation}\label{eq:applications.2D.rip}
    \e = \mu = \eta_o + \delta\eta\exp{\left(\pare{\frac{x-x_o-v\:t}{\sigma_x}}^2 + \pare{\frac{y-y_o}{\sigma_y}}^2\right)}
\end{equation}
where where $\eta_o\in\R$ is the background material parameter, $\delta\eta\in\R$ is the spacetime-variation's amplitude, $\square_o\in\R$ is the offset in the $x$ or $y$ direction; and $\sigma_i\in\R$ is the full-width half maximum in the corresponding direction. Specifically we set $\eta_o=1.5$, $\delta\eta=0.15$, $v=0.59$, $x_o=25$, $y_o=\frac{L_y}{2}$, $\sigma_x=5$ and $\sigma_y=25$.

Table \ref{table:applications.2D.errors} summarizes the errors and convergence rate for subsequent grid refinement at different grid sizes $h$ defined by the number of cell grids \mx.

\begin{table}[h!]
    \centering
    \begin{tabular}{r|r|rr}
        \hline
        \multicolumn{1}{c}{$mx/my$}  & \multicolumn{1}{c}{$h/mx$} & \multicolumn{1}{c}{$E_s(h)$} & \multicolumn{1}{c}{$p_s$} \\
        \hline
             128    & 1.978         &  1.981e-01    & 2.231 \\
             256    & 4.944e-01     &  4.222e-02    & 2.980 \\
             512    & 1.236e-01     &  5.349e-03    & 2.210 \\
             1024   & 3.090e-02     &  1.156e-03    & 2.008 \\
             2048   & 7.725e-03     &  2.875e-04    & 2.004 \\
            4096     & 1.931e-03    &  7.168e-05    & ---   \\
        \hline
    \end{tabular}
    \caption{Errors and rate of convergence for 2D electromagnetic wave propagation in a linear spacetime-varying medium.\label{table:applications.2D.errors}}
\end{table}

\begin{figure}[h!]
    \centering
    \includegraphics[width=0.6\textwidth]{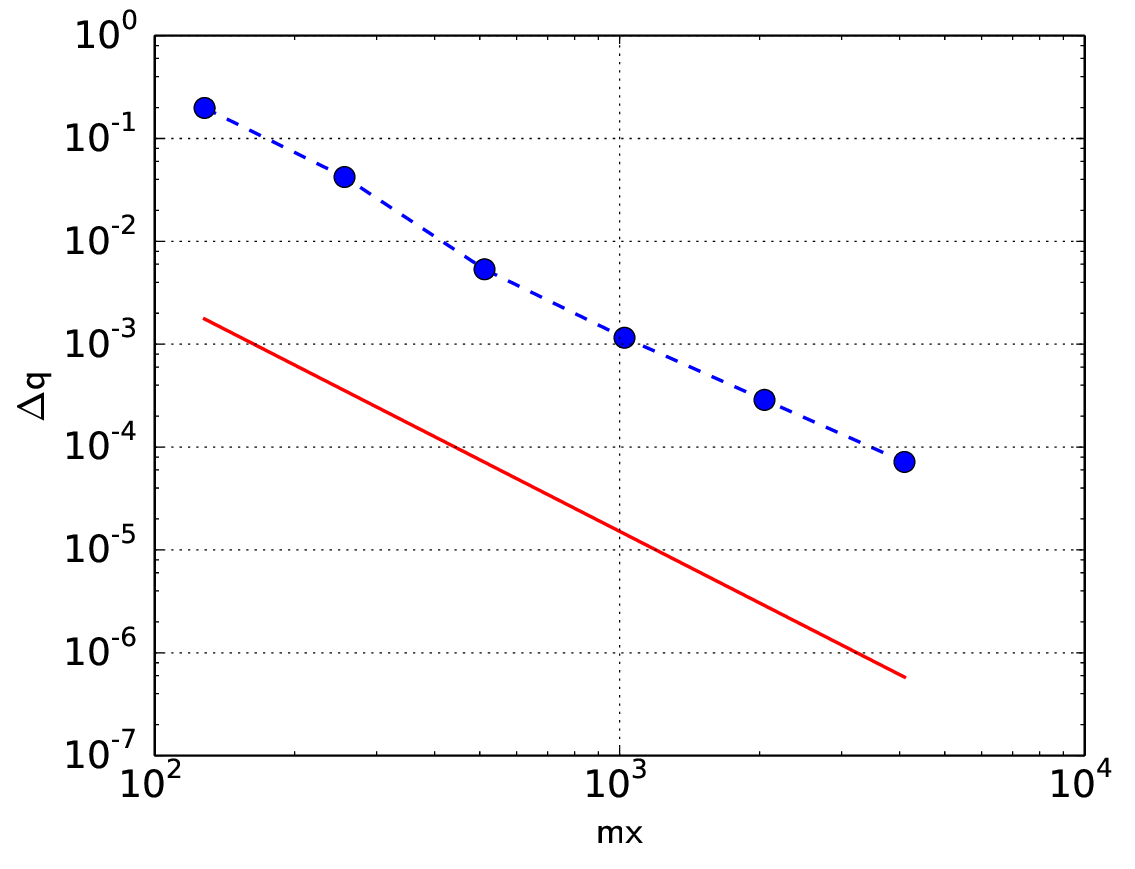}
    \caption{Convergence of the finite volume solution to subsequent refined finite volume solution, $E_s(h)$,  2D electromagnetic wave propagation in a linear spacetime-varying medium (blue-dashed); and liner fit with slope $p_s\approx -2.524$ (red).\label{fig:applications.2D.convergence}}
\end{figure}

Figure \ref{fig:applications.2D.convergence} shows $E_s(h)$. Again, using a least square linear fit, we find the convergence rate in the asymptotic region to be  $p_s\approx 2.524$.\par

\section{Summary}\label{sec:summary}
We have shown an approach to extend the highly accurate wave propagation algorithm of SharpClaw to model wave propagation in nonlinear and spacetime-varying media in one and two spatial dimensions. The scheme is second-order accurate, as demonstrated by the test results.

Using WENO and strong-stability-preserving time integration, high-order accurate results are obtained in one dimension even when the average value of the coefficients in each cell is second-order-accurate.

A drawback of our implementation is that  to achieve the high-order convergence in two dimensions, as observed in SharpClaw, the spacetime-varying coefficients need to be resolved with high accuracy, for example, using high-degree polynomial interpolation and adequate quadrature.

\bibliographystyle{siamplain}
\bibliography{library}

\end{document}